\documentclass[showpacs,preprintnumbers,amsmath,amssymb,floatfix]{revtex4}

\usepackage{color}   
\usepackage{graphicx}
\usepackage{dcolumn}
\usepackage{bm}

\begin{document}

\title{Klein-Gordon Equation in Hydrodynamical Form}

\author{C. Y. Wong}

\affiliation{Physics Division, Oak Ridge National Laboratory, Oak Ridge, 
TN\footnote{wongc@ornl.gov} 37831}

\date{\today}

\begin{abstract}

We follow and modify the Feshbach-Villars formalism by separating the
Klein-Gordon equation into two coupled time-dependent Schr\" odinger
equations for particle and antiparticle wave function components with
positive probability densities.  We find that the equation of motion
for the probability densities is in the form of relativistic
hydrodynamics where various forces have their classical counterparts,
with the additional element of the quantum stress tensor that depends
on the derivatives of the amplitude of the wave function.  We derive
the equation of motion for the Wigner function and we find that its
approximate classical weak-field limit coincides with the equation of
motion for the distribution function in the collisionless kinetic
theory.

\end{abstract}


\maketitle

\section{Introduction}

In relativistic wave mechanics, it is well known that the naive
probability density $\rho=2 {\rm Im} (\psi^* \partial_t \psi)$, as
constructed from the wave function $\psi$ of the Klein-Gordon
equation, is not necessarily a positive quantity.  The presence of a
negative probability density may appear to preclude its description in
relativistic hydrodynamics.  The origin of such pathology arises
because the wave equation is second order in the time-derivative, and
the corresponding probability density involves a time-derivative of
the wave function.  Dirac's resolution of the problem led to the
introduction of another wave equation, the Dirac equation, that is
first order in the time-derivative \cite{Dir28,Wei95}.  A parallel
resolution of the pathology was provided by Pauli and Weisskopf in
quantum field theory with the introduction of particles and
antiparticles, and the wave field $\psi$ is interpreted not as the
probability amplitude but as comprising of operators that create and
destroy particles in various modes \cite{Pau34}.

Wave mechanical interpretation of the wave field $\psi$ was however
reintroduced by Feshbach and Villars \cite{Fes58} by noting that the
Klein-Gordon equation actually contains both particle and antiparticle
degrees of freedom.  The particle-antiparticle separation can be
achieved by writing the Klein-Gordon and the Dirac equations as a set
of coupled time-dependent Schr\"odinger equations for the particle and
antiparticle wave function components with positive probability densities
\cite{Fes58}.

In many practical problems, as for example in the evolution of dense
matter with relativistic constituents in relativistic hydrodynamics
\cite{Lan53,Bjo83,Bay83,Oll92,hydro,Won94} or in the quarkonium
two-body problem \cite{Saz88,Cra88}, particles and antiparticles can be
approximately treated as two distinct types of interacting particles,
and they possess positive probability densities. For these problems,
we are motivated to follow the Feshbach-Villars formalism where the
particle and antiparticle probability densities can be positive
definite.

We wish however to modify the formulation of Feshbach and Villars
\cite{Fes58}.  In their formulation, the relativistic properties of
the momentum variables in the coupled Schr\"odinger equations are not
readily apparent and the Schr\"odinger equation for the antiparticle
wave function contains the operator $-({\bf p}-e{\bf A})^2/2m$ that
differs from the standard kinetic energy operator by a sign.  It will
be desirable to reformulate the problem so that the relativistic
properties of the momentum variables become more apparent and the
Schr\"odinger equations for the particle and antiparticle wave
functions contain kinetic energy operators with the same sign.  As
a result, the connection to relativistic hydrodynamics can be more
readily worked out.

Using the new set of coupled Schr\"odinger equations, we wish to
search for a hydrodynamical description for the evolution of
relativistic probability densities.  Hydrodynamics and quantum
mechanics have many elements in common, since the density fields and
the velocity fields are important dynamical variables in both
descriptions.  It is therefore well-known that the Schr\"odinger
equation can be cast into a hydrodynamical form for the evolution of
the probability density \cite{Mad26,Won76}.  Such a correspondence has
been utilized \cite{Won77} to form the foundation for an ``{\it a
posteriori}" theoretical support for the validity of treating a
nucleus as a liquid drop, as in Bohr and Wheeler \cite{Boh38}, and
treating the fission of a nucleus in liquid-drop hydrodynamics, as in
Hill and Wheeler \cite{Hil53}.  There are however important
differences associated with quantum effects that are absent in
classical hydrodynamics \cite{Bra72,Kan77,Won76,Won77}.  The deviation
of the classical hydrodynamical description from the quantum treatment
is embodied in the presence of the quantum stress tensor
$p_{ij}^{(q)}$ in quantum fluids \cite{Won76,Won77}. Quantum shell
effects manifest themselves as nuclear shell effects superimposed on a
smooth hydrodynamical liquid-drop background and they lead to the
intrinsic deformation in many nuclei \cite{Bra72}.

In the related area of hadron and nuclear collisions, a relativistic
hydrodynamical description of the collision process is a reasonable
concept, as pioneered by the work of Landau \cite{Lan53} and supported
by recent experimental findings \cite{Mur04,Ste05,Won08}.
Relativistic hydrodynamics has been applied to study the evolution of
a quark-gluon matter in the work of Bjorken \cite{Bjo83}, Baym
$et~al.$ \cite{Bay83}, Ollitraut \cite{Oll92}, and many others
\cite{hydro}.  They led to successful investigations on the dynamics
of matter with relativistic constituents in extreme conditions, as
occurs in relativistic heavy-ion collisions \cite{Mur04,Oll08}.  Wave
mechanical description of quantum systems in terms of probability
densities and probability currents contains the proper theoretical
ingredients appropriate for continuum hydrodynamics.  It will
therefore be of interest to generalize previous formulation of
hydrodynamics in the Schr\"odinger equation in Ref.\ \cite{Won76} to
Klein-Gordon and Dirac equations, in order to investigate how the wave
mechanics of Klein-Gordon and Dirac particles may be used to provide
the foundation for relativistic hydrodynamics of dense and compressed
systems with relativistic constituents.

As the foundation of hydrodynamics is usually presented within the
framework of the kinetic theory \cite{Hua63,deG80,Bai08,Gar08}, it is
instructive to investigate how the present quantum probability density
approach and the kinetic theory approach are connected.  We shall
study how the equation of motion for the Wigner function, derived from
the time-dependent Schr\"odinger equations representing the
Klein-Gordon equation, can be related to the equation of motion for
the phase space distribution function in the kinetic theory, in the
classical weak-field and collisionless limit.

This paper is organized as follows.  In Section II, we introduce the
particle and antiparticle wave functions to represent the Klein-Gordon
wave function and its time derivative.  The Klein-Gordon equation is
then separated into two coupled time-dependent Schr\"odinger
equations.  The particle and antiparticle probability densities obey
equations of continuity containing additional terms, but the total net
particle number remains a conserved quantity.  In Section III, we
examine the Euler equation for the motion of the probability fluid of
a particle or an antiparticle.  To provide insight into the dynamics,
we specialize to the simplified case when the particle-antiparticle
pair production can be suppressed.  The motion of the probability
fluids of particle and antiparticle obeys relativistic fluid dynamics
equations, and the quantum stress tensor provides part of the source
of the equation of state of relativistic matter.  In Section IV, we
examine the generalization to relativistic hydrodynamics for a simple
many-body system in the mean-field description.  In Section V, we
derive the equation of motion for the Wigner function and compare it
with the equation of motion for the distribution function in the
kinetic theory.  In Section VI, we present our summary and
discussions.

\section{Separation of Klein-Gordon equation into particle and
  antiparticle components}

In the present investigation, we are interested in the temporal
evolution of a single-particle boson state $\nu$ with an initial wave
function $\psi({\bf r},t)$ at $t=0$, in external fields that consist
of a scalar field ${\cal S}$ and a vector gauge field $(A^0,{\bf A} )$
with a coupling constant $e$.  The single-particle state is
characterized by a (net) particle number $n_\nu$ that is a conserved
quantity (see Eqs.\ (\ref{eq38}) and (\ref{na}) below).  The particle
number of a state is quantized; it has the value of $n_\nu=1$ when
$\nu$ is a particle state and $n_\nu=-1$ when $\nu$ is an antiparticle
state.  The evolution of the state $\nu$ is described by the
Klein-Gordon equation
\begin{eqnarray}
\label{kg}
(i\hbar\partial_{t}-e A^{0})^{2}\psi=[(\frac{\hbar}{i}\nabla-e {\bf A})^{2}+(m+{\cal S})^{2}]\psi.
\end{eqnarray}
To separate out the particle and antiparticle degrees of freedom, we
introduce the energy function $E$ and the auxiliary wave function
$\psi_{4}$.  For a given wave function $\psi$ for the state $\nu$ with a
charge $e_\nu =n_\nu e$, the energy function $E$ is the positive root of
the following  quadratic equation of $E$,
\begin{eqnarray}
\label{EE}
\int d{\bf r} ~~\psi^* \{  (E-e_\nu A^0)^2+[i\hbar \partial_t(E-e_\nu A^0)]  
-[(\frac{\hbar}{i}\nabla - e_\nu {\bf A} )^2 +(m+{\cal S})^2 ] \} \psi =0,
 \end{eqnarray}
which is obtained by taking the scalar product of the wave function
with the Klein-Gordon equation (\ref{kg}).  We may not know $
\partial_t E$ initially at $t=0$, but $E$ and $\partial_t E$ can
presumably be evaluated self-consistently and iteratively, when we
succeed in getting the equations (Eq.\ (\ref{sum}) below) that allow
us to advance the wave function to the next time step.  Following  Feshbach and Villars \cite{Fes58}, we introduce
the auxiliary wave function $\psi_4$ to represent  the time-derivative
of the wave function $\psi$,
\begin{eqnarray}
\label{KL1}
(i\hbar\partial_{t}-e A^{0})\psi=(E-e_\nu A^{0})\psi_4.
\end{eqnarray}
Our definition of $\psi_4$ in the above equation differs from that of
Feshbach and Villars \cite{Fes58}, where the right-hand side is given
as $m\psi_4$. The present definition exhibits better the relativistic
properties of the momentum variables and facilitates the
representation of the Klein-Gordon equation in relativistic
hydrodynamics.  The Klein-Gordon equation (\ref{kg}) becomes
\begin{eqnarray}
\label{KL2}
(i\hbar\partial_{t}-eA^{0})\psi_4
=\frac{1}{E-e_\nu A^{0}}\left 
\{(\frac{\hbar}{i}\nabla-e{\bf A})^{2}+(m+{\cal S})^{2}]\psi
-[i\hbar \partial_t(E-eA^0)] \psi_{4} \right \}.
\end{eqnarray}
We define the particle and antiparticle components 
$\chi_\pm$ of the state $\nu$  as linear combinations of $\psi$ and $\psi_4$,
\begin{eqnarray}
\label{def1}
\chi_{+}&=&\frac{1}{\sqrt{2}}[\psi+\psi_{4}]
\\
\chi_{-}&=&\frac{1}{\sqrt{2}}[\psi^{*}-\psi_{4}^{*}]
{\rm ~~~~or~~~}
\chi_{-}^{*}=\frac{1}{\sqrt{2}}[\psi\ -\psi_{4}].
\label{def2}
\end{eqnarray}
Our definition of $\chi_-$ in Eq. (\ref{def2}) contains a complex
conjugation that is different from the definition of $\chi_-$ of
Feshbach and Villars \cite{Fes58}.  Such a modification allows one to
obtain a Schr\"odinger equation for $\chi_-$ that contains the kinetic
energy operator $({\bf p}-e{\bf A})^2/2(E-e_\nu A^0)$, instead of a
kinetic energy operator with the opposite sign in Feshbach and Villars
\cite{Fes58}.  In terms of the wave functions $\chi_+$ and $\chi_-$ as
defined in Eqs.\ (\ref{def1}) and (\ref{def2}), equations (\ref{KL1})
and (\ref{KL2}) become
\begin{eqnarray}
\label{sum}
(i\hbar\partial_{t}-e_{\pm}A^{0})\chi_{\pm}
&=&\frac{1}{2(E-e_\nu A^0)}\left \{(\frac{\hbar}{i}\nabla-e_{\pm}{\bf A})^{2}+(m+{\cal S})^{2}+[(E-e_\nu A^0)^{2}-i\hbar \partial_t(E-e_\nu A^0)] \right \}\chi_{\pm}
\nonumber\\
&+&
\frac{1}{2(E-e_\nu A^0)}\left \{(\frac{\hbar}{i}\nabla-e_{\pm}{\bf A})^{2}
+(m+{\cal S})^{2}-[(E-e_\nu A^0)^{2}-i\hbar \partial_t(E-e_\nu A^0)]\right  \}\chi_{\mp}^*,
\end{eqnarray}
where $e_\pm=\pm e$. Thus we obtain the central result that the
Klein-Gordon equation can be reduced to a set of two time-dependent
Schr\"odinger equations in which the particle and the antiparticle
appear as distinct types of particles with opposite charges $e_\pm$
with the particle wave function $\chi_+$ and the antiparticle wave
function $\chi_-$.  A general solution of the Klein-Gordon equation
contains two components that can be represented by a column vector,
\begin{eqnarray}
\label{Psi}
\Psi=
\left ( 
\begin{matrix} 
 \chi_+\cr  
\chi_-  \cr 
\end{matrix}
 \right ).
\end{eqnarray}
Both $\chi_+$ and $\chi_-$ have positive norms, $|\chi_\pm|^2$, which
can be interpreted as the probability densities of particles and
antiparticles respectively, as in hydrodynamics.  

Note that our denominator in Eqs. (\ref{sum}) differs from the
denominator of $m$ in the formulation of Feshbach and Villars
\cite{Fes58}.  The relativistic properties of the momentum variable
are more apparent in Eq.\ (\ref{sum}).  These relativistic properties
will facilitate the subsequent representation of the Klein-Gordon
equation in relativistic hydrodynamical form.

The second term of the new equation (\ref{sum}) represents the
particle-antiparticle coupling and pair production.  It is
proportional to $((\hbar/i)\nabla-e_{\pm}{\bf A})^{2} +(m+{\cal
S})^{2}-[(E-e_\nu A^0)^{2}-i\hbar \partial_t(E-e_\nu A^0)] $ and
involves essentially the deviation of $E^2$ from ${\bf p}^2+m^2$ that
increases with the strength of the interaction.  The coupling is
inversely proportional to the energy (or mass) of the particle.  The
particle-antiparticle coupling and the rate of pair production are
large, when the strength of the interaction relative to the rest mass
of the particle is large \cite{Sch51,Wan88,Won94}.  The
particle-antiparticle coupling is small when the strength of the
interaction relative to the energy (or mass) of the particle is small.

For stationary states $\chi_\pm$ with $e_\nu=n_\nu e=\pm e$ in static
external fields, Eq.\ (\ref{sum}) gives
\begin{eqnarray}
\left \{ (E-e_\pm A^0)^2 -(\frac{\hbar}{i}\nabla -e_\pm{\bf A})^2
  -(m+{\cal S})^2\right \} \chi_\pm
= 
\left \{ - (E-e_\pm A^0)^2+(\frac{\hbar}{i}\nabla -e_\pm{\bf A})^2
  +(m+{\cal S})^2  \right \} \chi_\mp^*.
\end{eqnarray}
If the right-hand side representing the pair production coupling can be
neglected, the above equation is just the time-independent
Klein-Gordon equation for a stationary state of a particle (or an
antiparticle) in external fields.  Equation (\ref{sum}) has the
correct limit for stationary states in static external fields in the
absence of pair productions.

In a static case with time-independent external fields, the energy
function $E$ is a constant of motion, although the wave functions
$\chi_\pm$ and the associated probability fluids in the static fields
can still evolve dynamically in space-time.  With time-dependent
external fields, the energy of the single-particle state changes
because energy may be supplied or removed by the external fields.  We
envisage conceptually that a two-component wave function $\Psi({\bf
  r},t)=(\chi_+({\bf r},t),\chi_-({\bf r},t))$ for the state $\nu$ with
$e_\nu=n_\nu e$ is initially known as $t=0$.  The knowledge of the
initial wave functions allows us to obtain $E$ from Eq.\ (\ref{EE})
and to construct $\chi_+$ and $\chi_-$ at the next time step with the
help of Eq.\ (\ref{def1}) and (\ref{def2}).  In the initial evaluation
of $E$ and in the subsequent stepwise increment of the wave function
in time, we need the quantity $\partial_t E(t)$, which may be
self-consistently determined by an iterative procedure.

After the wave function $\Psi$ has been advanced to the next time
step, we can evaluate $E(t)$ using Eq. (\ref{EE}) that is a quadratic
equation in $E$.  It contains both a positive-$E$ and a negative-$E$
solution.  We shall limit our attention only on the positive-$E$
solution so that we can speak of both particle and antiparticles
states of positive energies.  This stepwise increment allows the
determination of the evolution of the state $\Psi({\bf r},t)$ and
$E(t)$ as a function of time.

To see how the probability fluids behave in space and time, we write
the wave functions $\chi_\pm$ in terms of the amplitude and phase
functions,
\begin{eqnarray}
\chi_\pm({\bf r},t)=\phi_\pm ({\bf r},t) e^{iS_\pm({\bf r},t)/\hbar-i\Omega(t)}.
\end{eqnarray}
We construct $\chi_\pm^*$$\times$(\ref{sum})-$\chi_\pm$$\times$(\ref{sum})$^*$.
After some manipulations, we find
\begin{eqnarray}
\label{con}
\partial_{t}[(E-e_\nu A^0) \phi_\pm^2]
+ \nabla\cdot [\phi_\pm^2(\nabla S_\pm -e_\pm {\bf A}) ]
=
X_\pm,
\end{eqnarray}
where
\begin{eqnarray}
2X_\pm &=&\{ \chi_\pm^*(\frac{\hbar}{i}\nabla-e_{\pm}{\bf A})^{2}\chi_\mp^*
- \chi_\pm(\frac{\hbar}{-i}\nabla-e_{\pm}{\bf A})^{2}\chi_\mp \}
\nonumber\\
& & +[(m+{\cal S})^2+(E-e_\nu A)^2] 
(\chi_\pm^*\chi_\mp^*-\chi_\pm \chi_\mp)
\nonumber\\
& & +
[i\hbar \partial_t(E-e_\nu A^0)] (\chi_\pm^*\chi_\mp^*+\chi_\pm\chi_\mp).
\end{eqnarray}
As the quantities $X_+$ and $X_-$ are not generally a full divergence,
the total number of particles and antiparticles in the two components
are not conserved.  However, the difference of the particle number and
antiparticle numbers of the two components satisfies the equation
\begin{eqnarray}
\partial_{t}[(E-e_\nu A^0) (\phi_+^2-\phi_-^2)]
+ \nabla\cdot [\phi_+^2(\nabla S_+ -e_+{\bf A})] 
- \nabla\cdot [\phi_-^2(\nabla S_- -e_- {\bf A})] 
=
X_+-X_-.
\end{eqnarray}
But, $X_+$-$X_-$ is a complete divergence,
\begin{eqnarray}
\label{XX}
X_+-X_-
=& -&
\nabla \cdot (\chi_+^* \nabla \chi_-^*-\chi_-^* \nabla \chi_+^*)/2
+\nabla \cdot (\chi_+ \nabla \chi_- - \chi_- \nabla \chi_+)/2
\nonumber\\
&-&
\nabla \cdot [e_+ A^0 ( \chi_+^*\chi_-^*+\chi_+ \chi_-)/i].
\end{eqnarray}
Therefore, the quantity 
\begin{eqnarray}
\label{eq38}
n_{\rm particle}=\int d{\bf r} \frac {E-e_\nu A^0}{m} (\phi_+^2-\phi_-^2)
=\int d{\bf r} \frac {E-e_\nu A^0}{m} (|\chi_+|^2-|\chi_-|^2)
\end{eqnarray}
is a conserved quantity because the right-hand side of the equation
(\ref{XX}) is a complete divergence.  The additional number of
particles produced is equal to the additional number of antiparticles
produced.  The equal increase in particle and antiparticle numbers
associated with $X_+$ and $X_-$ represents the occurrence of
particle-antiparticle pair production.

A single-particle solution with a $n_{\rm particle}=n_\nu=\pm 1$ is
one in which $|\chi_\pm|^2 \gg |\chi_\mp|^2$ and can be normalized to be
\begin{eqnarray}
\label{na}
\int d^3r \frac{E-e_\nu A^0}{m} [|\chi_\pm|^2 -|\chi_\mp|^2]=1 {\rm ~~ for~a~particle~state~with~}n_{\rm particle}=\pm 1.
 \end{eqnarray}

\section{Klein-Gordon equation in hydrodynamical form}

To obtain the Euler equation, we construct $\chi_\pm^*$$\times$(\ref{sum})+$\chi_\pm$$\times$(\ref{sum})$^*$.
Putting all terms together, we get
\begin{eqnarray}
\label{euler}
& &\phi_\pm^2(-2\partial_t S_\pm-2e_\pm A^0 ) 
\nonumber\\
&=&
\frac{1}{2(E-e_\nu A^0)} \biggl  \{
-[2\phi_\pm \nabla^2 \phi_\pm-2\phi_\pm^2(\nabla S_\pm
-e_\pm {\bf A} )^2 ]
\nonumber\\
& &
+[(m+{\cal S})^2+(E-e_\nu A^0)^2] 2\phi_\pm^2 
\nonumber\\
& &+
\chi_\pm^*(\frac{\hbar}{i}\nabla-e_{\pm}{\bf A})^{2} \chi_\mp^*
+\chi_\pm(\frac{\hbar}{-i}\nabla-e_{\pm}{\bf A})^{2}\chi_\mp 
\nonumber\\
& &+
[(m+{\cal S})^2-(E-e_\nu A)^2] 
(\chi_\pm^*\chi_\mp^*+\chi_\pm \chi_\mp)
\nonumber\\
& &+
[i\hbar \partial_t(E-e_\nu A^0)] (\chi_\pm^*\chi_\mp^*-\chi_\pm\chi_\mp)
\biggr \}.
\end{eqnarray}

In the present hydrodynamical description, we investigate the
evolution of the probability densities of particles and antiparticles
in a single-particle system with a definite particle number $n_\nu$ in
external fields.  The rate of the particle-antiparticle pair
production depends on the strength of the external interaction
relative to the particle rest mass.  In the case of a strong
interaction and large pair production probabilities, a
hydrodynamical description will be complicated as it will involve a 
component that describes the pair production process.

A hydrodynamical description will be appropriate after the active pair
production stage has passed and the expansion of the system is driven
by a slowly varying external field or a mean field.  It is this type
of motion for which we wish to provide a hydrodynamical description.
This is the case when the interaction is present but not large
compared to the rest mass, so that the probability of pair production
is small, and can be suppressed in the lowest-order approximation.
This is equivalent to the case of a ``simple fluid" in relativistic
hydrodynamics, in which the chemical composition of the fluid ceases to
change \cite{Lan86}. In such circumstances, one can speak of a single-particle
system with a definite particle number $n_\nu$, which can take on
$n_\nu=1,-1$ values.  We shall now consider this simplifying case with 
suppressed pair production by discarding the last four terms in the
curly bracket on the right hand-side of Eq.\ (\ref{euler}).  After
dividing by $-2\phi_\pm^2$, the equation for the phase function
$S_\pm$ for this simplified case is
 \begin{eqnarray}
(\partial_t S_\pm+e_\pm A^0 ) 
&=&
\frac{1}{2(E-e_\nu A^0)} \biggl  \{
[(\nabla^2 \phi_\pm)/\phi_\pm-(\nabla S_\pm
-e_\pm {\bf A} )^2 ]
\nonumber\\
& &
-(m+{\cal S})^2-(E-e_\nu A^0)^2  \biggl  \}.
\end{eqnarray}
For this case with suppressed pair production, $e_\nu=n_\nu e = e_\pm$.
We take the gradient $\nabla_i$ of the above for $i=1,2,3$, and
multiply by $\phi_\pm^2(E-e_\pm A^0)$.  We obtain
\begin{eqnarray}
& &(E-e_\pm A^0)\phi_\pm^2 \partial_t (\nabla_i S_\pm-e_\pm A^i) 
\nonumber\\
& =& \biggl \{ \phi_\pm^2 \nabla_i
[(\nabla^2 \phi_\pm)/2\phi_\pm]
-  \sum_{j=1}^3 \phi_\pm^2 (\nabla_j S_\pm-e_\pm  A^j ) \nabla_j (\nabla_i S_\pm-e_\pm A^i ) 
\nonumber\\
& &
-(m+{\cal S}) \nabla_i {\cal S}-\sum_{j=1}^3 \phi_\pm^2 (\nabla_j S_\pm-e_\pm  A^j )e_\pm F^{ij}
\biggr \}- (E-e_\pm A^0)\phi_\pm^2e_\pm F^{0i}
\nonumber\\ 
& &
+\frac{e_\pm \nabla_i  A^0\phi_\pm^2}{2(E-e_\pm A^0)} \biggl  \{
(\nabla^2 \phi_\pm)/\phi_\pm-(\nabla S_\pm-e_\pm {\bf A} )^2 ]
-(m+{\cal S})^2+(E-e_\pm A^0)^2\biggr \}.
 \nonumber\\ 
\end{eqnarray}
Using the equation of continuity for this simplified case without
pair production, we obtain
\begin{eqnarray}
& &
\partial_t\left [ \frac{(E-e_\pm A^0)\phi_\pm^2(\nabla_i S_\pm-e_\pm A^i)}{m+{\cal S}}\right ] +\sum_{j=1}^3 \nabla_j
\left [\frac {\phi_\pm^2 (\nabla_j S_\pm-e_\pm  A^j )  (\nabla_i S_\pm-e_\pm A^i ) }{m+{\cal S}} \right ]
\nonumber\\
& =&
-\frac{m}{m+{\cal S}}\sum_{j=1}^3 \nabla_j  p_{ij}^{(q)}
 - \phi_\pm^2\nabla_i {\cal S}
+\frac{1}{m+{\cal S}} \biggl \{ 
- (E-e_\pm A^0)\phi_\pm^2e_\pm F^{0i}
-\sum_{j=1}^3 \phi_\pm^2 (\nabla_j S_\pm-e_\pm  A^j )e_\pm F^{ij}
\biggr \}
\nonumber\\ 
& &
+\frac{e_\pm \nabla_i  A^0\phi_\pm^2}{2(E-e_\pm A^0)(m+{\cal S})} 
\biggl  \{
(\nabla^2 \phi_\pm)/\phi_\pm-(\nabla S_\pm-e_\pm {\bf A} )^2 ]
-(m+{\cal S})^2+(E-e_\pm A^0)^2\biggr \}
\nonumber\\ &  &
-(E-e_\pm A^0)\phi_\pm^2(\nabla_i S_\pm-e_\pm A^i) 
\frac{\partial_t {\cal S}} {(m+{\cal S})^2}
-\sum_{j=1}^3 
\left [\frac {\phi_\pm^2 (\nabla_j S_\pm-e_\pm  A^j )  (\nabla_i S_\pm-e_\pm A^i ) }{m+{\cal S}} \right ]
\frac{\nabla_j {\cal S}} {(m+{\cal S})^2}. 
\end{eqnarray}
We can identity the fluid energy density $\epsilon_\pm$ as 
\begin{eqnarray}
\epsilon_\pm=(m+{\cal S})\phi_\pm^2,
\end{eqnarray}
as it corresponds to the fluid energy density for the fluid element at rest.
The fluid element is characterized by a relativistic 4-velocity $u^\mu$.  We
can identify
\begin{eqnarray}
u_\pm^0&=&\frac{E-e_\pm A^0}{m+{\cal S}}
\nonumber\\
u_\pm^i&=&\frac{\nabla_i S_\pm-e_\pm A^i}{m+{\cal S}}, {\rm ~~~~for ~} i=1,2,3,
\end{eqnarray}
which obeys $(u_\pm^0)^2 - ({\bf u}_\pm)^2=1$ in the absence of the
quantum effects.  We can then write an equation of motion for the
probability densities in terms of the hydrodynamical equation
\begin{eqnarray}
\label{hyd}
  & &\partial_t ( \epsilon_\pm u_\pm^0 u_\pm^i)
  + \sum_{j=1}^3 \nabla_j  \epsilon_\pm u_\pm^i u_\pm^j
  +\frac{m}{m+{\cal S}}   \sum_{j=1}^3 \nabla_j p_{\pm ij}^{(q)}
  \nonumber\\
  & =&
  - \phi_\pm^2\nabla_i {\cal S}
  +\frac{1}{m+{\cal S}} \biggl \{ -
  (E-e_\pm A^0)\phi_\pm^2e_\pm F^{0i}
  -\sum_{j=1}^3 \phi_\pm^2 (\nabla_j S_\pm-e_\pm  A^j )e_\pm F^{ij}
  \biggr \}
  \nonumber\\ 
  & &
  +\frac{e_\pm \nabla_i  A^0\phi_\pm^2}{2(E-e_\pm A^0)(m+{\cal S})} 
  \biggl  \{
  (\nabla^2 \phi_\pm)/\phi_\pm-(\nabla S_\pm-e_\pm {\bf A} )^2 ]
  -(m+{\cal S})^2+(E-e_\pm A^0)^2\biggr \}
  \nonumber\\ &  &
  -(E-e_\pm A^0)\phi_\pm^2(\nabla_i S_\pm-e_\pm A^i) 
  \frac{\partial_t {\cal S}} {(m+{\cal S})^2}
  -\sum_{j=1}^3 
  \left [\frac {\phi_\pm^2 (\nabla_j S_\pm-e_\pm  A^j )  (\nabla_i S_\pm-e_\pm A^i ) }{m+{\cal S}} \right ]
  \frac{\nabla_j {\cal S}} {(m+{\cal S})^2},
  \nonumber\\ 
\end{eqnarray}
where $i,j=1,2,3.$ This is the Klein-Gordon equation for the particle
and antiparticle probability densities in hydrodynamical form.  The
first two terms on the left-hand side correspond to $\partial_\mu
T_\pm^{\mu i}$, with the energy momentum tensor of the probability
fluid $T_\pm^{\mu i}=\epsilon_\pm u_\pm^\mu u_\pm^i$, for
$\mu=0,1,2,3$.  The third term on the left-hand side is the quantum
stress tensor arising from the spatial variation of the amplitude of
the single-particle wave function \cite{Won76},
\begin{eqnarray}
p_{\pm\, ij}^{(q)} =-\frac{\hbar^2}{4m}\nabla^2\phi_\pm^2 \delta_{ij}
+\frac{\hbar^2}{m}\nabla_i \phi_\pm \nabla_j \phi_\pm.
\end{eqnarray}
  In Eq.\ (\ref{hyd})
there is no thermal stress tensor contribution in Eq.\ (\ref{hyd}) for
a single-particle.  As indicated in \cite{Won76}, a thermal stress
tensor $p_{ij}^{(t)}$ will however arise from different velocity
fields when there are many particles in a many-body
system.  The first two terms on the right-hand side contain forces
coming from the scalar interaction, the electric field $F^{0i}$ and
the magnetic field $F^{ij}$, as in classical formulations.  The
third term on the right-hand side is the relativistic correction to the
time-like part of the vector interaction, and the last two terms
represent relativistic corrections associated with the spatial and
temporal variation of the scalar interaction.  Thus the dynamics of
the probability fluid obeys an equation similar to the hydrodynamical
equation, with forces on fluid elements arising from what one
expects in classical considerations.   The additional element is the
presence of the quantum stress tensor $p_{ij}^{(q)}$ that is proportional to $\hbar^2$ and arises from
the quantum nature of the fluid.

While we have examined the hydrodynamical form when the
particle-antiparticle pair production has been suppressed, the
particle-antiparticle pair production can be included in future
studies, at the expenses of increasing greatly the complexity of
the simple picture we have obtained.

\section{Application to a Many-Body System in 
the Mean-Field Description}

The external mean field we have been studying can come internally from
the single-particle state probability density as in the
Gross-Pitaevskii equation \cite{Gro61,Gro63,Pit61,Pit03} or from the
self-consistent mean-field in a many-body
system \cite{Won76,Won77,Bon74}.  A many-body system in the
time-dependent mean-field description consists of a collection of
independent particles moving in self-consistent mean-fields generated
by all other particles \cite{Bon74,Won76,Won77}.  Each single-particle
state $\psi_{a \nu}$ is characterized by a state label $a$, particle
type $\nu$, energy $e_\nu$, and occupation number $n_{a\nu}$.  For
simplicity, we consider the case in which the mean-field potential
arises from a scalar two-body interaction $v_s({\bf r}_1, {\bf r}_2)$
and a time-like vector interaction $v_0({\bf r}_1, {\bf r}_2)$.  We
further neglect the last three terms on the right-hand side of Eq.\
(\ref{hyd}) which represent higher-order relativistic corrections.
The equation of motion for the energy density $\epsilon_{a\nu}$ and
velocity fields $u_{a\nu}^i$ for $i=1,2,3$ and $\nu=\pm$, in the
single particle state $a$ and particle type $\nu$, is then
\begin{eqnarray}
\label{hyd24}
  & &\partial_t ( \epsilon_{a\nu} u_{a\nu}^0 u_{a\nu}^i)
  + \sum_{j=1}^3 \nabla_j  \epsilon_{a\nu} u_{a\nu}^i u_{a\nu}^j
  +\frac{m}{m+{\cal S}}   \sum_{j=1}^3 \nabla_j p_{ (a\nu)ij}^{(q)}
  \nonumber\\
  & =&
  - \phi_{a\nu}^2\nabla_i {\cal S}
  +\frac{E-e_{a\nu} A_\pm^0}{m+{\cal S}} 
  \phi_{a\nu}^2e_{a\nu} \frac{\partial A^0}{\partial x^i} ,
\end{eqnarray}
where, in the frame with the fluid element at rest,
\begin{eqnarray}
{\cal S}({\bf r},t) =\int d^3{\bf r}_2~ n({\bf r_2},t) v_s({\bf r},{\bf r}_2),
\end{eqnarray} 
\begin{eqnarray}
A^0({\bf r},t) =\int d^3{\bf r}_2~ \biggl \{ n_+({\bf r_2},t) e_+
+n_-({\bf r_2},t) e_-\biggr \}
 v_0({\bf r},{\bf r}_2),
\end{eqnarray}
\begin{eqnarray}
n_\nu=\sum_a n_{a\nu}\phi_{a\nu}^2, ~~ 
{\rm ~~and~~}
n=n_++n_-.
\end{eqnarray}
We consider a strongly interacting system in which the number of
 particles and antiparticles are equal so that $n_+({\bf
 r_2})=n_-({\bf r_2})$ and $n_+({\bf r_2}) e_+ +n_-({\bf r_2}) e_-$ is
 zero.  Then the contribution from the second term on the right-hand
 side of Eq.\ (\ref{hyd24}) is zero.  Multiplying Eq.\ (\ref{hyd24})
 by $n_{a\nu}$ and summing over $\{a,\nu\}$, we get
\begin{eqnarray}
& & \partial_t (\sum_{a\nu}n_{a\nu}\epsilon_{a\nu} u_{a\nu}^0 u_{a\nu}^i )
  + \sum_{j=1}^3 \nabla_j   (\sum_{a\nu}n_{a\nu}\epsilon_{a\nu} u_{a\nu}^i u_{a\nu}^j)
\nonumber\\
& &
  +\frac{m}{m+{\cal S}}   \sum_{j=1}^3  \nabla_j ( \sum_{a\nu}n_{a\nu}p_{(a\nu) ij}^{(q)})
  + (\sum_{a\nu}n_{a\nu}\phi_{a\nu}^2) \nabla_i {\cal S} 
=0.
\end{eqnarray}
We define the total energy density $\epsilon$ by
\begin{eqnarray}
\sum_{a\nu} n_{a\nu} \epsilon_{a\nu}=\epsilon,
\end{eqnarray}
and the average  4-velocity $u$ by 
\begin{eqnarray}
u = {\sum_{a\nu}  n_{a\nu} \epsilon_{a\nu} u_{a\nu}}
       /{\epsilon}.
\end{eqnarray}
We can introduce the thermal stress tensor $p_{ij}^{(t)}$ for
$\{i,j\}=1,2,3$ as the correlation of the deviations of the
single-particle velocity fields from the average
\begin{eqnarray}
\sum_{a\nu} n_{a\nu} \epsilon_{a\nu} (u_{a\nu}^i-u^i) (u_{a\nu}^j -u^j)
\equiv p_{ij}^{(t)}.
\end{eqnarray}
For the case with the suppression of pair production, we obtained the equation of hydrodynamics 
\begin{eqnarray}
\partial_t (\epsilon u^0 u^i )
+ \sum_{j=1}^3 \left \{ \nabla_j \left  (\epsilon u^i u^j +p_{ij}^{(t)}+p_{ij}^{(v)}\right )
+\frac{m}{m+{\cal S}}  \nabla_j  p_{ ij}^{(q)} \right \}
=0 ,
\end{eqnarray}
where the total quantum stress tensor is 
\begin{eqnarray}
p_{ ij}^{(q)} =-\frac{\hbar^2}{4m}\nabla^2\sum_{a\nu} n_{a\nu} \phi_{a\nu}^2 \delta_{ij}
+\frac{\hbar^2}{m}\sum_{a\nu}n_{a\nu}\nabla_i \phi_{a\nu} \nabla_j \phi_{a\nu},
\end{eqnarray}
and the pressure due to the interaction $p_{ij}^{(v)}$ is 
\begin{eqnarray}
\frac{\partial}{\partial x^j} p_{ij}^{(v)}({\bf r},t)=n({\bf r},t)\nabla_i {\cal S}({\bf r},t)=
n({\bf r},t) \frac{\partial}{\partial x^j}
\int d^3 {\bf r}_2 n({\bf r}_2,t) v_s({\bf r},{\bf r}_2).
\end{eqnarray}
The mean-field stress tensor $p_{ij}^{(v)}$ can also be given as
\begin{eqnarray}
 p_{ij}^{(v)}=
\left \{ n \frac{\partial (W^{(v)} n) }{\partial n}
- W^{(v)} n \right \}\delta_{ij},
\end{eqnarray}
where $W^{(v)}$ is the energy per particle arising from the mean-field
interaction.  As an illustrative example, we can consider a
density-dependent two-body interaction
\begin{eqnarray} 
v_s( {\bf r},{\bf r}_2)=[a_2+a_3 n(({\bf r}+{\bf r}_2)/2)]
\delta({\bf r}-{\bf r}_2).
\end{eqnarray}
The contribution of the mean-field interaction to the stress tensor is then
\begin{eqnarray}
 p_{ij}^{(v)}=
 \frac{1}{2}[a_2+2a_3 n({\bf r})] {n^2} \delta_{ij},
\end{eqnarray}
whose magnitude increases with the density and the strengths of the
interaction.

The quantum stress tensor $p_{ij}^{(q)}$ and the thermal stress tensor
$p_{ij}^{(t)}$ can take on different values, depending on the
occupation numbers $n_{a\nu}$ of the single-particle states that
determine the degree of thermal equilibrium of the system.  The
quantum stress tensor depends on the amplitudes of the wave functions
while the thermal stress tensor depends on the phases of the wave
functions and the deviation of the velocity fields from the mean
velocities.  The quantum stress tensor is less sensitive to the degree
of thermalization as compared to the thermal stress tensor.  In the
time-dependent mean-field description, the motion of each particle
state can be individually followed \cite{Bon74}.  The occupation
numbers $n_{a\nu}$ of the single-particle states will remain
unchanged, if there are no additional residual interaction between the
single particles due to residual interactions.  When particle residual
interactions are allowed as in the extended time-dependent mean-field
approximation \cite{Won78}, the occupation numbers will change and
will approach an equilibrium distribution as time proceeds.

We note that the total pressure arises from many sources. We come to
the observation that in situations when $|p_{ij}^{(q)}+p_{ij}^{(v)}|
\gg p_{ij}^{(t)}$ for a strongly-coupled system at low and moderate
temperatures, there can be situations when the system behaves
quasi-hydrodynamically, even though the state of the system has not
yet reached thermal equilibrium.  In this case, the hydrodynamical
state is maintained mainly by the quantum stress tensor and the strong
mean fields.

What we have discussed in this Section is only a theoretical framework
that exhibits clearly the different sources of stress tensors. To
study specifically the dynamics of the quark-gluon plasma, for
example, it will be necessary to investigation the specific nature of
different constituents and their interactions. Nevertheless, the
general roles played by the different components of stress tensors can
still be a useful reminder on the importance of the quantum and
mean-field stress tensors in the strongly-coupled regime, at
temperature just above the transition temperature $T_c$.

\section{Connection to the Kinetic Theory}
\label{section5}

The foundation of hydrodynamics is usually presented within the
framework of kinetic theory, in which particles and antiparticles are
described as distinct constituents and their interactions are
weak \cite{Hua63,deG80,Bai08,Gar08}.  In such a description, particles
are considered to be approximately on-the-mass-shell, and their
inter-particle collisions lead to thermalization.  The state of the
system for particles or antiparticles of type $\nu$ is described by a
distribution function $f_\nu({\bf r},{\bf p};t)$ in phase space.
Successive gradient expansions of small deviations from the
equilibrium distribution lead to various approximations of the
transport coefficients. The time-dependencies of the moments of
various kinematic operators lead to hydrodynamical equations.  The
near-mass-shell condition restricts its application to systems with
weak interactions that can be taken as perturbations in quantum field
theory.

For a dense and strongly interacting system, such as a nucleus or a
strongly-coupled quark-gluon plasma, a reasonable description of a
non-equilibrium system can be formulated in a different approach, the
quantum probability density approach considered here.  In this
approach, constituents of the quantum system move in the strong mean
fields generated by all other particles.  Each particle executes its
single-particle motion in the time-dependent mean field, and the
residue interaction between particles lead to ``collisions'' that
change the single-particle state occupation numbers $n_{a\nu}$.  These
residual-interaction collisions occur in such a way that the single
particle occupation numbers approaches a thermal distribution as a
function of time
\cite{Won78}. 

In the quantum probability density approach, as the particles resides
in a strong field, they are off-the-mass shell and their energies
depend on their local potential.  The dynamics of each single particle
state is described by a wave function with a positive probability
density and a probability current.  The equations for the total
probability density and probability current are analogous to the
relativistic hydrodynamical equations.  The stress tensors in such a
description arises from many components: the quantum stress tensor,
the thermal stress tensor, and the mean-field interaction stress
tensor.  The behavior of the dynamics in such a quantum description
need not be the same as classical hydrodynamics because the
constitutive equations relating various stress components with
densities and other attributes may respond differently to the evolving
dynamics \cite{Won77a}.  We mention earlier that in situations in which
the quantum stress tensor and interaction stress tensor dominate over
the thermal stress tensor, the degree of thermalization may not be
important in the hydrodynamical behavior of the system.  In the other
extreme when the stress tensor arises predominantly from the thermal
stress tensor, the dynamics will then depend on the degree of thermal
equilibrium.

Although the quantum probability density approach is particularly
appropriate for cases of strongly interacting quantum systems, it also
has a well-defined classical and weak-coupling limit that should
coincide with the kinetic theory approach.  It is therefore
instructive to investigate how the two approaches are connected by
studying how the equation of motion for the Wigner function, derived
from the time-dependent Schr\"odinger equations representing the
Klein-Gordon equation, can be related to the equation of motion for
the phase space distribution function in the kinetic theory, in the
classical weak-field and collisionless limit.

Both in the present quantum probability density approach and the
kinetic theory approach, the application to relativistic hydrodynamics
are confronted with the problem of pair production in which the
interactions will lead to spontaneous production of
particle-antiparticle pairs \cite{Sch51,Wan88}.  The standard
hydrodynamics is one in which pair-production probability is
suppressed.  The equation of motion of the single-particle state of
particle type $\nu$ is then obtained from Eq. (\ref{sum}) by
neglecting the second term on the right-hand side.  We further
consider the motion to be sufficiently slow so that we can neglect the
term $\partial_t(E-e_\nu A^0)$.  The equation of motion for the
single-particle state is then
\begin{eqnarray}
\label{sch}
(i\hbar\partial_{t}-e_{\nu}A^{0})\chi_{\nu}
&=&\frac{1}{2(E-e_\nu A^0)}\left \{(\frac{\hbar}{i}\nabla-e_{\nu}{\bf A})^{2}+(m+{\cal S})^{2}+[(E-e_\nu A^0)^{2}] \right \}\chi_{\nu},
\end{eqnarray}
which can be written as
\begin{eqnarray}
i\hbar\partial_{t}\chi_{\nu}
&=&\frac{1}{2(E-e_\nu A^0)}\left \{(\frac{\hbar}{i}\nabla-e_{\nu}{\bf A})^{2}\right \}\chi_{\nu}
+V(r) \chi_{\nu},
\end{eqnarray}
where
\begin{eqnarray}
V(r)
&=&\left \{\frac{(m+{\cal S})^{2} }{2(E-e_\nu A^0)}+\frac{E+e_{\nu}A^{0}}{2} \right \}.
\end{eqnarray}
From this time-dependent Schr\"odinger equation, we wish to obtain the
corresponding equation of motion for the single-particle Wigner
function.  The construction of a gauge invariant Wigner function from
the density matrix in the presence of an external gauge field $(A^0,
{\bf A})$ has been examined by many authors
\cite{Irv65,Ser86,Bia91,Bia93,Lev94,Lev01,Hoe02,Var03,Lev09,Haa10},
and we can follow similar procedures. 
We construct $ \chi_{\nu}^\dagger ({\bf r}_1, t)
\chi_{\nu}({\bf r}_2, t) $ and introduce $({\bf r}_1+{\bf r}_2)/2={\bf
  r}$, and ${\bf r}_1-{\bf r}_2={\bf s}$.  We define the
three-dimensional gauge-invariant single-particle Wigner function for
particle type $\nu$ as
\begin{eqnarray}
\label{wig}
 f_\nu({\bf r},{\bf p}; t)
&=&\int d{\bf s} e^{i {\bf p} \cdot {\bf s}}
\{{\cal F}_I\}^{-1}
 \chi_{\nu}^\dagger({\bf r}+{\bf s}/2, t )\chi_{\nu} ({\bf r}-{\bf s}/2, t)
\end{eqnarray}
where the momentum ${\bf p}$ represents the kinetic
momentum \cite{Lev94,Lev01}, and ${\cal F}_I$ is the well-known gauge
factor introduced first by Schwinger \cite{Sch62},
\begin{eqnarray}
{\cal F}_{I}
&=&
\exp \{- ie_\nu \int_{{\bf r}-{\bf s}/2}^{{\bf r}+{\bf s}/2}
{\bf A}({\bf s}')\cdot  d{\bf s}' \}.
\end{eqnarray} 
The inverse transform of Eq.\ (\ref{wig}) is then
\begin{eqnarray}
\label{inv}
\chi_{\nu}^\dagger({\bf r}+{\bf s}/2, t )
\chi_{\nu} ({\bf r}-{\bf s}/2, t)
= \int \frac{d{\bf p}}{2\pi\hbar^3}
e^{-i {\bf p}  \cdot {\bf s}}
\exp \{- ie_\nu \int_{{\bf r}-{\bf s}/2}^{{\bf r}+{\bf s}/2}
{\bf A}({\bf s}')\cdot  d{\bf s}' \}
 f_\nu({\bf r},{\bf p}; t).
\end{eqnarray}
From the Schr\"odinger equation (\ref{sch}), we construct
$\chi_{\nu}^\dagger ({\bf r}_1, t)
i\hbar\partial_{t}
\chi_{\nu}({\bf r}_2, t)$ and
$
\chi_{\nu} ({\bf r}_2, t)
[-i\hbar\partial_{t}]
\chi_{\nu}^\dagger({\bf r}_1, t)$,
subtract the two, and we get
\begin{eqnarray}
i\hbar\partial_{t}
\label{full}
[\chi_{\nu}^\dagger({\bf r}_1, t)\chi_{\nu} ({\bf r}_2, t)]
&=&
\biggl  [\left \{\frac{1}{2[E-e_\nu A^0({\bf r}_2)]} \left (\frac{\hbar}{i}\nabla_{{\bf r}_2}-e_{\nu}{\bf A}({\bf r}_2)\right )^{2}+V({\bf r}_2)\right \}
\nonumber\\
& &-
\left \{\frac{1}{2[E-e_\nu { A}^0({\bf r}_1)]} \left (\frac{\hbar}{-i}\nabla_{{\bf r}_1}-e_{\nu}{\bf A}({\bf r}_1)\right)^{2}+V({\bf r}_1)\right \}  \biggr ]
\chi_{\nu}^\dagger ({\bf r}_1, t) \chi_{\nu}({\bf r}_2, t).
\end{eqnarray}
The equation of motion for the Wigner function $f_\nu({\bf r},{\bf
p})$ can be obtained by substituting Eq.\ (\ref{inv}) into the above
equation.  The terms involving $V({\bf r})$ give the result
\begin{eqnarray}
\label{vr}
\biggl [ V({\bf r}_2)-V({\bf r}_1)   \biggr ]\chi_{\nu}^\dagger({\bf r}+{\bf s}/2, t )
\chi_{\nu} ({\bf r}-{\bf s}/2, t)
= \int \frac{d{\bf p}}{2\pi\hbar^3}
e^{-i {\bf p}  \cdot {\bf s}}
{\cal F}_{I}
 \frac{2}{\hbar}\sin \biggl \{ \frac{\hbar}{2} \nabla_{\bf p}^f  \cdot \nabla _{\bf r}^V\biggr \} V({\bf r})
f_\nu({\bf r},{\bf p}; t).
\end{eqnarray}
where $ \nabla_{\bf p}^{f}$ applies only on $f_\nu$ and $\nabla _{\bf
  r}^{V}$ applies only on $V({\bf r})$. This indicates that the
quantum equation of motion for the Wigner function contains
transcendental functions of the operators $\hbar \nabla_{\bf p} \cdot
\nabla _{\bf r}$ applying on the potential and the Wigner function.
The expansion of the sine function will lead to a power series in
$\hbar$.  We see here that the equation of motion for the distribution
function in kinetic theory corresponds only the lowest order
approximation in such an expansion.  We note also that the Wigner
function is in general not identical to the distribution function
because it can take on negative values \cite{Wig32,Won03}.  In the
classical limit, the Wigner function can be confined to be positive
and be identified with the phase space distribution function.  To make
connections with the classical kinetic theory approach, we shall take
this limit of $\hbar \to 0$.  The above equation is then approximated
as
\begin{eqnarray}
\label{vr}
\biggl [ V({\bf r}_2)-V({\bf r}_1)  \biggr ]\chi_{\nu}^\dagger({\bf r}+{\bf s}/2, t )
\chi_{\nu} ({\bf r}-{\bf s}/2, t)
= \int \frac{d{\bf p}}{2\pi\hbar^3}
e^{-i {\bf p}  \cdot {\bf s}}
{\cal F}_{I}
 [\nabla _{\bf r}V({\bf r})]
 \cdot  \nabla_{\bf p}
f_\nu({\bf r},{\bf p}; t).
\end{eqnarray} 
We shall take the weak-field limit so that we keep only terms first order in the external fields.    Then the quantity $\nabla_{\bf r} V({\bf r})$
is 
\begin{eqnarray}
\nabla_{\bf r} V(r)&=&\left \{
\frac {(m+{\cal S}) \nabla {\cal S} }{(E-e_\nu A^0)}
+\frac{(m+{\cal S})^{2} e_\nu \nabla A^0}{2(E-e_\nu A^0)^2}+\frac{\nabla e_{\nu}A^{0}}{2} +..\right \},
\end{eqnarray}
which goes to 
$\nabla {\cal S}+\nabla e_{\nu}A^{0}$  in the non-relativistic limit of $E\to m$.

To evaluate the other terms, we note that the variation of the gauge
factor ${\cal F}_I$ arising from the variation of the end point ${\bf
r}_2$ is given by
\begin{eqnarray}
\delta_{{\bf r}_2} e^{-ie_\nu 
\int_{{\bf r}_2}^{{\bf r}_1}
{\bf A}({\bf s}')\cdot d{\bf s}'}
=-i e^{-ie_\nu \int_{{\bf r}_2}^{{\bf r}_1}
{\bf A}({\bf s}')\cdot d{\bf s}'}
~e_\nu \delta_{{\bf r}_2} 
[\int_{{\bf r}_2}^{{\bf r}_1} {\bf A}({\bf s}')\cdot d{\bf s}'].
\end{eqnarray}
The difference of the path integrals when one of the end points is
varied from ${\bf r}_2$ to ${\bf r}_2 + \delta {\bf r}_2$ can be
turned into a loop integral by noting that
\begin{eqnarray}
\delta_{{\bf r}_2} 
[\int_{{\bf r}_2}^{{\bf r}_1} {\bf A}({\bf s}')\cdot d{\bf s}']
&=&
\left [\int_{{{\bf r}_2+\delta {\bf r}_2}}^{{\bf r}_1}- \int_{{\bf r}_2}^{{\bf r}_1} \right ]  {\bf A}({\bf s}')\cdot d{\bf s}' 
\nonumber\\
&=&
\left [
   \oint_{{\bf r}_2 \to {{\bf r}_2+\delta {\bf r}_2}
              \to  {\bf r}_1 \to {\bf r}_2 }
-  \int_{{\bf r}_2}^{{{\bf r}_2+\delta {\bf r}_2}}
 \right ]  {\bf A}({\bf s}')\cdot d{\bf s}' 
\nonumber\\
&=&
   \biggl [\oint_{{\bf r}_2 \to {{\bf r}_2+\delta {\bf r}_2}
              \to  {\bf r}_1 \to {\bf r}_2 }
{\bf A}({\bf s}')\cdot d{\bf s}' \biggr ]
-    {\bf A}({\bf r}_2)\cdot \delta{\bf r}_2.
\end{eqnarray}
By Stokes' theorem, we can carry out the loop integral over a
triangular area encircled by the loop and we have
\begin{eqnarray}
   \oint_{{\bf r}_2 \to {{\bf r}_2+\delta {\bf r}_2}
              \to  {\bf r}_1 \to {\bf r}_2 }
{\bf A}({\bf s}')\cdot d{\bf s}' 
&=&
\int_{\rm area~ encircled~by~loop} 
\nabla \times {\bf A} \cdot [\delta {\bf r}_2\times d{\bf s}']
\nonumber\\
&\sim&
{\bf B}({\bf r}) \cdot \frac {[\delta {\bf r}_2\times {\bf s}]}{2}
\sim 
\frac{1}{2}\delta {\bf r}_2 \cdot [ {\bf s}\times {\bf B}({\bf r})].
\label{bbb}
\end{eqnarray}
Here a Taylor expansion of the ${\bf B}({\bf r}+{\bf s}')$ field ${\bf
B}({\bf r}+{\bf s}')$ in powers of ${\bf s}'$ in the loop integral
will lead to terms in power of $\hbar$ \cite{Lev94,Lev01}.  Taking the
${\bf B}({\bf r})$ field to be located at ${\bf r}$ in Eq.\
(\ref{bbb}) represents the lowest-order $\hbar \to 0$ approximation in
the expansion of ${\bf B}({\bf r}+{\bf s}')$.  As $\delta_{{\bf
r}_2} \phi =\nabla_{{\bf r}_2} \phi \cdot
\delta {\bf r}_2$, we obtain
\begin{eqnarray}
\nabla_{{\bf r}_2} 
e^{-ie_\nu \int_{{\bf r}_2}^{{\bf r}_1}
 {\bf A}({\bf s}')\cdot d{\bf s}'}
&=&
e^{-i e_\nu \int_{{\bf r}_2}^{{\bf r}_1}
{\bf A}({\bf s}')\cdot d{\bf s}'}
i \left \{ e_\nu  {\bf A} ({\bf r}_2)
- \frac{1}{2}
 [ {\bf s}\times e_\nu {\bf B}({\bf r})]
 \right \},
\end{eqnarray}
and
\begin{eqnarray}
\left ( \frac{\hbar}{i}\nabla_{{\bf r}_2}- e_\nu {\bf A}({\bf r}_2)\right  )
e^{-i e_\nu \int_{{\bf r}_2}^{{\bf r}_1}
 {\bf A}({\bf s}')\cdot d{\bf s}'}
&=&
e^{-i e_\nu  \int_{{\bf r}_2}^{{\bf r}_1}
{\bf A}({\bf s}')\cdot d{\bf s}'}
\left \{  -\frac{1}{2} [ {\bf s}\times e_\nu {\bf B}({\bf r})]
 \right \}.
\label{eq32}
\end{eqnarray}
Therefore, we obtain
\begin{eqnarray}
&  &
\biggl \{(\frac{\hbar}{i}\nabla_{{\bf r}_2}-e_{\nu}{\bf A}({\bf r}_2))^2 -\{(\frac{\hbar}{-i}\nabla_{{\bf r}_1}-e_{\nu}{\bf A}({\bf r}_2))^2\biggr \}
\biggl[ e^{-ie_\nu \int_{{\bf r}_2}^{{\bf r}_1}
 {\bf A}({\bf s}')\cdot d{\bf s}'}
e^{ -i {\bf p} \cdot {\bf s} }
  f_\nu({\bf r},{\bf p})\biggr]
\nonumber\\
&=&
\left \{
 e^{-i e_\nu \int_{{\bf r}_2}^{{\bf r}_1}
 {\bf A}({\bf s}')\cdot d{\bf s}' } \right \}
\left \{ 
\left [\frac{\hbar}{i}\nabla_{{\bf r}_2}
- \frac{1}{2}
 {\bf s}\times e_\nu {\bf B}({\bf r})
\right ]^2
-\left [\frac{\hbar}{-i}\nabla_{{\bf r}_1}
+ \frac{1}{2}
 {\bf s}\times e_\nu {\bf B}({\bf r})
\right ]^2\right \} 
e^{ -i {\bf p} \cdot {\bf s} }
  f_\nu({\bf r},{\bf p})
\nonumber\\
\end{eqnarray}
After some manipulation for the remaining terms,  we get
\begin{eqnarray}
\label{result}
& &\int \frac{d{\bf p}}{2\pi\hbar^3}
{\cal F}_{I}
e^{-i {\bf p} \cdot {\bf s}}  \biggl  \{ 
i \hbar \partial_t   f_\nu({\bf r},{\bf p})
+ i \frac{ {\bf p} }{E} 
\cdot \nabla_{\bf r}  f_\nu({\bf r},{\bf p})
+i (-\nabla_{\bf r} V +\frac{{\bf p}}{E}\times e_\nu {\bf B}({\bf r}))
\cdot \nabla_{\bf p} f_\nu({\bf r},{\bf p})
 \biggr \}=0,
\end{eqnarray}
which is satisfied if
\begin{eqnarray}
\label{finalf}
\partial_t   f_\nu({\bf r},{\bf p})
+  \frac{ {\bf p}}{E} 
\cdot \nabla_{\bf r}  f_\nu({\bf r},{\bf p})
+ (-\nabla_{\bf r} V +\frac{{\bf p}}{E}\times e_\nu {\bf B}({\bf r}))
\cdot \nabla_{\bf p} f_\nu({\bf r},{\bf p})
=0.
\end{eqnarray}
This is just the equation of motion for the distribution function in kinetic theory for a collisionless fluid in weak fields \cite{Ich73}.

Equation (\ref{finalf}) is the equation of motion for the Wigner
function of a single-particle state $\nu$ in external scalar and gauge
fields.  For a many-body system with different states $a$ and particle
types $\nu$, the Wigner function is
\begin{eqnarray}
f({\bf r},{\bf p}) = \sum_{ a\nu} n_{ a\nu}  f_{a\nu} ({\bf r},{\bf p}).
\end{eqnarray}
  The equation of motion for the total distribution function $f({\bf
r},{\bf p})$ in the collisionless limit will be in the same form as
Eq. (\ref{finalf}) with $f_\nu({\bf r},{\bf p})$ in Eq. (\ref{finalf})
replaced by $f({\bf r},{\bf p})$. The presence of other particles in
different single-particle states allows one use the mean-fields as the
external fields and to introduce the collision term by considering
residual interactions between particles in different $a\nu$ states.

The exercise in this Section indicates that although the quantum
probability density approach is particularly appropriate for cases of
strongly interacting quantum systems, it also has a well-defined
classical weak-coupling limit that coincides with the kinetic theory,
from which hydrodynamics equations can also be formulated.

\section{Summary and Discussions}

We have generalized the formulation of Feshbach and Villars to write
the Klein-Gordon equation as a set of two coupled time-dependent
Schr\"odinger equations, for the particle and antiparticles components
of the wave function.  We have improved upon the formulation of
Feshbach and Villars in this re-examination.  In our new set of
coupled time-dependent Schr\"odinger equations (\ref{sum}) the
particle and the antiparticle components are better separated, and the
kinetic energy operator in the equation for the antiparticle component
is in the proper form of $({\bf p}-e{\bf A})^2/2(E-e_\nu A^0)$, with a
positive sign. The relativistic properties of the momentum variables
in the coupled time-dependent Schr\"odinger equations are more
apparent and their connection to relativistic hydrodynamics can be
easily established.

We introduce amplitude and phase functions to cast the time-dependent
Schr\"odinger equations into hydrodynamical form.  We find that the
equation of motion of the probability fluid of Klein-Gordon particles
or antiparticles can be written in the form of relativistic
hydrodynamics, with an additional quantum stress tensor.  The other
components of the hydrodynamical equation have their classical
counterparts.

For simplicity, we have suppressed the pair-production degree of
freedom in the present investigation.  The pair production is however
a quantum phenomenon that can be studied by using the Klein-Gordon
equation, as was carried out in Ref.\ \cite{Wan88} to examine the
Schwinger mechanism. The presence of the pair-production mechanism is
a new feature in the hydrodynamical evolution.  Since energy and
momentum is diverted into pair production, the pair-production
corresponds to a dissipative process, and will contribute to the
viscosity of the fluid.  Future investigations to include this
pair-production in relativistic hydrodynamics will be of great
interest.

As both the Schr\"odinger equation and the Klein-Gordon equation can
be cast into a hydrodynamical form, one may inquire whether the Dirac
equation can also be written in hydrodynamical form. It is well known
that the Dirac equation can be reduced to a Klein-Gordon equation,
with additional terms involving the spin and particle-antiparticle
degrees of freedom.  For a Dirac particle in an external field we have
\begin{eqnarray}
\{\gamma^\nu (i \partial_\nu - e A_\nu) - (m+{\cal S})\}
\psi = 0.
\end{eqnarray}
Upon multiplying this on the left by $\gamma^\nu (i \partial_\nu - e A_\nu) +
(m+{\cal S})$, we obtain
\begin{eqnarray}
\{ (i \partial_\nu - e A_\nu)^2 - (m+{\cal S})^2
-i {\bf \alpha}  \cdot  e {\bf E} + {\bf \sigma}\cdot e {\bf B}({\bf r}) 
- i [\gamma^\nu \partial_\nu {\cal S}]\} 
\psi = 0,
\end{eqnarray}
which is the Klein-Gordon equation with additional interactions.
Thus, the Dirac equation can be likewise cast into a hydrodynamical
form, following the procedures outlined in the present manuscript.

We have thus far discussed particles and antiparticles with opposite
charges interacting in a gauge field. For the case of neutral
particles, there is no interaction with the gauge field characterized
by the charges $e_\pm$.  However, relativistic doubling of states
occurs and there are particles and antiparticles.  How these neutral
particles should be treated will depend on their interaction with the
external field. If the interactions of the neutral particle and
antiparticle with the external fields are identical, it will not be
necessary to distinguish between a particle and an antiparticle.  It
then becomes possible to construct a simplified theory in which only
one degree of freedom enters (say, the particle's), with a neutral
antiparticle taken to be identical to its corresponding neutral
particle.  On the other hand, if their interactions with the external
fields are different, the two degrees of freedom are distinct. One can
then introduce additional quantum numbers to distinguish particles and
antiparticles, and the present investigation containing different
charges (or quantum numbers) for the new types of interaction will
apply.

In the dynamics of a single-particle state in an external field, the
external field can come internally from the single-particle state
probability density as in the Gross-Pitaevskii
equation \cite{Gro61,Gro63,Pit61,Pit03} or from the self-consistent
mean-field in a many-body system \cite{Won76,Won77}.  A many-body
system consists of a collection of various particles in their
individual single-particle states.  For example, we can examine a
many-body system in the time-dependent Hartree approximation in which
particle and antiparticles move in a time-dependent self-consistent
mean-field generated by all other particles.  The motion of each
particle can be individually followed, as in the time-dependent
mean-field approximation in a nuclear system \cite{Bon74}.  The
occupation numbers of these single-particle states will remain
unchanged, if there are no additional residual interaction between the
single particles.  When particle residual interactions are allowed as
in the extended time-dependent mean-field approximation \cite{Won78},
the occupation numbers will change and will approach an equilibrium
distribution.  As the fluid density is cumulative in nature the
dynamics of the many-body system will rely on the cumulative effects
from particles in individual states.  Thus, there will be a total
quantum stress tensor that is an important part of the equation of
state of the many-body system. There will also be contributions from
the mean fields to the equation of states of the relativistic fluid as
in the nuclear fluid \cite{Won77}.  The present investigation of
Klein-Gordon single-particle states should be useful in the study of
the dynamics of a many-body system using relativistic hydrodynamics.

The foundation of hydrodynamics is usually presented within the
framework of kinetic theory, we can compare the present quantum
probability density approach to the usual kinetic theory approach by
determining the equation of motion for the Wigner function using the
time-dependent Schro\"dinger equation of the Klein-Gordon equation.
Our comparison indicates that the equation of motion for the
distribution function in an external field in the kinetic theory can
be obtained from the quantum probability density approach in the limit
of $\hbar \to 0$ and weak fields.  Thus, the quantum probability
approach extends the range of hydrodynamics applications to situations
where quantum effects and/or strong interactions are important.

\vspace {0.5cm}
\centerline{\bf Acknowledgment}
\vskip .5cm
The author would like to thank Profs.\ H. W. Crater,
S. S. Willenbrock, L. S. Garcia-Colin, and T. Barnes for helpful
discussions and communications. The research was sponsored by the
Office of Nuclear Physics, U.S. Department of Energy.

\end{document}